\documentclass[final,5p,times,twocolumn]{elsarticle}

\usepackage{amssymb}
\usepackage{amsthm}
\usepackage{amsmath}
\usepackage{xcolor}

\usepackage{lineno}

\journal{Physics Letter B}

\begin{document}

\begin{frontmatter}



\title{Implications for Dark Matter Direct Detection in the Presence of \\ LIGO-Motivated Primordial Black Holes}


\author[1]{Mark P. Hertzberg}
\ead{mark.hertzberg@tufts.edu}
\author[2,3,4]{Enrico D. Schiappacasse}
\ead{enrico.e.schiappacasse@jyu.fi}
\author[4]{Tsutomu T. Yanagida}
\ead{tsutomu.tyanagida@ipmu.jp}
\address[1]{Institute of Cosmology, Department of Physics and Astronomy, Tufts University, Medford, MA 02155, USA}
\address[2]{Department of Physics, P.O.Box 35 (YFL), FIN-40014 University of Jyv$\ddot{a}$skyl$\ddot{a}$, Finland}
\address[3]{Helsinki Institute of Physics, P.O. Box 64, FIN-00014 University of Helsinki, Finland}
\address[4]{Tsung-Dao Lee Institute $\&$ School of Physics and Astronomy, Shanghai Jiao Tong University, 
 200240 Shanghai, China}

\begin{abstract}
We discuss formation of dark matter (DM) mini-halos around primordial black holes (PBHs) and its implication on DM direct detection experiments, including axion searches. Motivated by LIGO observations, we consider $f_{\textrm{DM}} \simeq 0.01$ as the fraction of DM in PBHs with masses $10 M_{\odot} - 70 M_{\odot}$.  In this case, we expect the presence of dressed PBHs after Milky Way halo formation with mini-halo masses peaked around $M_{\textrm{halo}} \sim (50-55) M_{\textrm{PBH}}$. We analyze the effect of tidal forces acting on dressed PBHs within the Milky Way galaxy. In the solar neighborhood, the mini-halos are resistant against tidal disruption  from the mean-field potential of the galaxy and encounters with stars, but they undergo a small level of disruption caused by disk shocking. The presence of mini-halos around LIGO-motivated PBHs today could reduce by half the local dark matter background. High-resolution simulations are encouraged. If the proposed scenario is realized, chances of direct detection of DM would decrease.
\end{abstract}




\end{frontmatter}

%\linenumbers


%\begin{linenumbers}
\section{Introduction}
\label{intro}

In the present study, we will explore the consequences on direct detection of dark matter on earth, if there are primordial black holes (PBHs). This is motivated by observations of binary black hole mergers by LIGO. The key idea is that the PBHs may accumulate dark matter mini-halos, which may impact the local dark matter distribution.

Recently, the idea of PBHs~\cite{Hawking:1971ei, Carr:1974nx, Carr:1975qj, Kawasaki:1997ju, GarciaBellido:1996qt, Khlopov:2008qy}  has been strongly revitalized since the first detection
of two merging black holes by the LIGO-Virgo collaboration~\cite{Abbott:2016blz}. This first detection                                  (GW150914) corresponds to the merger of two black holes with masses $\sim30M_{\odot}$.
Assuming all black hole binaries relevant to the LIGO observation share the same mass and other physical parameters, the estimation for the merger event rate is $2 - 53$ \,$\textrm{Gpc}^{-3}$ yr$^{-1}$~\cite{Abbott2016}. The coalescence of PBH binaries can successfully explain this merger rate if PBHs constitute a small fraction of dark matter~\cite{Nakamura:1997sm, Ioka:1998nz, Sasaki:2016jop}. 

If the merger of PBH binaries is actually the source of gravitational wave events detected by LIGO-Virgo collaboration, estimates of the PBH merger rate can be translated into an (potential) upper bound on $f_{\textrm{DM}} \equiv \Omega_{\textrm{PBH}}/\Omega_{\textrm{DM}}$. This bound is stronger than most of current observational constraints~\cite{Tisserand:2006zx, Allsman:2000kg, Brandt:2016aco, 2014ApJ...790..159M, Ali-Haimoud:2016mbv} for  $10-300 M_{\odot}\,\textrm{PBH}$~\footnote{We do not consider CMB constraints reported by Ricotti et al. (2008)~\cite{Ricotti:2007au}, which overestimate the effects of PBHs on CMB observables (see~\cite{Ali-Haimoud:2016mbv}).}. If PBHs comprise a significant fraction of dark matter, they would produce a merger rate larger than those observed by LIGO~\cite{Sasaki:2016jop,  Eroshenko:2016hmn, Ali-Haimoud:2017rtz, Raidal:2017mfl}. 

Assuming an extended PBH mass function and the possibility of a clustered spatial distribution, the fraction of dark matter in PBHs is estimated in~\cite{Raidal:2017mfl}. The merger of PBH binaries can explain the merger rate inferred by LIGO~\cite{Abbott:2017vtc} if $f_{\textrm{DM}} = 4.5\times10^{-3} - 2.4\times10^{-2}$ for 3$0M_{\odot}\, \textrm{PBHs}$ and a lognormal mass function. 

By considering tidal forces over the PBH binary coming from all remaining PBHs and standard large-scale adiabatic perturbations, potential upper bounds on $f_{\textrm{DM}}$ as a function of  PBH masses are discussed in~\cite{Ali-Haimoud:2017rtz}.  LIGO O1  run~\cite{Abbott:2017iws, Abbott:2016drs, TheLIGOScientific:2016htt} would constrain  $\sim 10\,M_{\odot} - 300\,M_{\odot}$ PBHs to be a fraction of dark matter no more than $1\%$.
In particular, potential upper bounds for $10\,M_{\odot}$ and $300\,M_{\odot}$ PBHs are reported to be $f_{\textrm{DM}} \lesssim 8 \times 10^{-3}$ and  $9 \times 10^{-3}$, respectively. 

Analytical estimates suggest that most of the PBH binaries evolve without disruption until merger~\cite{Ali-Haimoud:2017rtz}.  This is expected for the case $f_{\textrm{DM}} \ll 1$, but as the fraction of dark matter in PBHs increases, disruption of binaries should become more significant. A robust calculation of the binary merger rate needs to include a suitable suppression in the merger rate due to disruption coming from a third PBH close and/or dense PBH clusters (see~\cite{Vaskonen:2019jpv} for a discussion and estimates).

The presence of dark mini-halos around PBHs would only have a mild effect on the binary merger rate. Numerical computations reported in~\cite{Kavanagh:2018ggo} showed that semi-major axis and the eccentricity of binaries will both decrease as a result of the dynamical friction and tidal forces exerted by dark mini-halos. For the mass range $10\,M_{\odot} - 300\,M_{\odot}$, the presence of dark mini-halos constrains $f_{\textrm{DM}}$ to be no more than $4\times 10^{-3}$~\cite{Kavanagh:2018ggo}. However, this constraint does not consider disruption of dressed PBH binaries after their formation coming from other isolated PBHs or clustered spatial distribution. This disruption should be enhanced in the presence of dark mini-halos.
 
Current constraints on $f_{\textrm{DM}}$ derived from gravitational wave observations include several theoretical uncertainties.    For our purposes, motivated by LIGO-Virgo detections and observational constraints, we take the conservative value $f_{\textrm{DM}} \simeq 0.01$ for the characteristic fraction of dark matter in  $\sim 10\,M_{\odot}$ - $70\,M_{\odot}$ PBHs~\footnote{Here we take into account CMB-anisotropy constraints from photoionization of the local gas from PBH radiation~\cite{Ali-Haimoud:2016mbv}.}.

To find  the second dark matter component to accompany PBHs, we need to look for beyond the Standard Model. By considering shortcomings in this model, the weakly interacting massive particle (WIMP) and the axion~\cite{Peccei:1977hh, PhysRevD.16.1791, Weinberg:1977ma, Wilczek:1977pj, Marsh:2015xka,  Kiritsis:2014yqa} are the strongest candidates. However,  the WIMP and PBHs cannot coexist unless the dark matter fraction in PBHs is highly suppressed~\cite{Lacki:2010zf, Adamek:2019gns, Bertone:2019vsk}. Assuming  self-annihilation of WIMPs during the late Universe into standard model particles (e.g. there is no significant WIMP-antiWIMP asymmetry, or dominant annihilation into hidden sectors), inner regions of WIMPS mini-halos around PBHs would undergo self-annihilation becoming strong sources of gamma rays and neutrinos\,\footnote{WIMPs with masses $\gtrsim 100\, \textrm{TeV}$ would evade this kind of constraint (private communication with S. Shirai).}.  So arguably, if the LIGO events are due to PBHs, the most serious dark matter candidate to accompany PBH dark matter may be the axion.

Even though we are primarily motivated by the axion in this work, we are not excluding other possible dark matter candidates.
Since formation of dark mini-halos around PBHs holds for any dark matter candidate\,\footnote{However, we expect accretion will not be efficient for ultra-light axions (or ultra-light scalar dark matter particles) when their De Broglie wavelength is comparable with or larger than the halo radius.}, conclusions of this letter are extensive to axion-like particles or light bosons having a symmetry breaking energy scale large enough to avoid sizeable self-annihilation in the most inner shells of the mini-halo. 
%\end{linenumbers}

%\begin{linenumbers}
PBHs are local overdensities in the dark matter distribution and they act as seeds for the formation of dark matter structures.  
Analytical and numerical calculations~\cite{Bertschinger:1985pd, Mack:2006gz} show dark mini-halos around PBHs~\footnote{A PBH with a mini-halo is sometimes called in the literature as dressed PBH.} mainly grow during the matter-dominated era reaching up to  $\sim10^2 M_{\textrm{PBH}}$ in units of the central PBH mass. In detail, we have~\cite{Mack:2006gz, Ricotti:2007au}
%\end{linenumbers}
\begin{align}
M_{\textrm{halo}} (z) &= 3\left(\frac{1000}{1+z}\right) M_{\textrm{PBH}}\,,\label{mh}\\
R_{\textrm{halo}} (z) = 0.019\, &\textrm{pc} \left( \frac{M_{\textrm{halo}}}{M_{\odot}} \right)^{1/3} \left( \frac{1000}{1+z} \right)\,,\label{rh}
\end{align}
%\begin{linenumbers}
where $R_{\textrm{halo}}$ is about one-third of the turnaround radius. Equations~(\ref{mh},\ref{rh}) assume a growth of the dark mini-halo in absence of external tidal forces. This assumption only holds  before the epoch of formation of the first galaxies at redshift $z \sim 30$.  On the other hand, as dressed PBHs begin to form a significant part of the dark matter, the halo growth rate begins to decrease~\cite{Mack:2006gz}. Thus, $M_{\textrm{halo}}(z\simeq 30)/M_{\textrm{PBH}}\simeq 10^2$  needs to be seen as an upperbound for the maximum mass of a PBH mini-halo at that time. 

From the above, we need to take seriously the possibility that nowadays a significant part of dark matter is part of dark mini-halos surrounding PBHs inside galactic halos. If this is the case, then the capability of earth based direct detection experiments, such as the ability of ADMX to detect axions~\cite{RYBKA201414, Stern:2014wma}, would be compromised.

Even though we are mainly interested in LIGO-motivated PBHs with masses $10 M_{\odot} - 70 M_{\odot}$, our warning is extensive to PBHs with larger, or smaller, masses.  If we consider only collisional ionization of the background gas in CMB-anisotropy constraints, e.g. the radiation coming from PBHs is not able to ionize the local gas~\cite{Ali-Haimoud:2016mbv}, one may extend the mass range of PBHs of our interest to $\sim 10M_{\odot} - 300 M_{\odot}$ for $f_{\textrm{DM}} \simeq 0.01$. In addition, primordial black holes which are formed with a mass $M_{\textrm{PBH}} \gtrsim 1.6 \times 10^{-17}\,M_{\odot}$ do not evaporate~\footnote{PBHs lighter than $3\times10^{-19}\,M_{\odot}$ evaporate within the present age of the Universe.} but begin to form compact dark-mini halos by accreting the surrounding dark matter. Since they may compose $1\%$ of dark matter without tension with observational constraints~\cite{Carr:2009jm, Barnacka:2012, Graham:2015apa, Niikura:2017zjd, Inomata:2017okj}, they may accrete in the form of mini-halos a significant fraction of dark matter background.

As far as we know, this is the first time that LIGO-motivated PBHs with dark mini-halos are proposed as a factor to be considered with respect to direct detection of dark matter on the Earth.   

\section{Evolution of dressed PBHs}
\label{evolve}

In the case of PBHs correspond to an initial fraction of dark matter $f_{\textrm{DM}} \ll 1$, numerical simulations reported in~\cite{Mack:2006gz} suggest that mini-halos around PBHs reach $\simeq10^2 M_{\textrm{PBH}}$ at around $z \sim 30$. However, if larger values of $f_{\textrm{DM}}$ are considered, then dressed PBHs during the accretion process will eventually become a sizeable part of the dark matter abundance.  At that time, the accretion rate becomes slower due to the decrease in the dark matter density background. In particular, for initial values of $f_{\textrm{DM}} \simeq 10^{-3} - 10^{-2}$, we have  $M_{\textrm{halo}} \simeq 10^2 M_{\textrm{PBH}} -  50 M_{\textrm{PBH}}$ at around $z \sim 30$ in the PBH mass range of our interest~\cite{Mack:2006gz}.

In our specific case, the fraction of dark matter in LIGO-motivated PBHs range as $\mathcal{O}(10^{-3}) \lesssim f_{\textrm{DM}} \lesssim \mathcal{O}(10^{-2})$.  Define $f^{z\sim30}_{\textrm{halo}}$ as the fraction of dark matter in the form of dressed PBHs at the time of first galaxies formation. Since $M_{\textrm{halo}} \gg M_{\textrm{PBH}}$ in the case of our interest, we have $f^{z\sim30}_{\textrm{halo}} \equiv f_{\textrm{DM}}[(M_{\textrm{halo}} + M_{\textrm{PBH}})/M_{\textrm{PBH}}]  \simeq f_{\textrm{DM}}(M_{\textrm{halo}}/M_{\textrm{PBH}})$. The maximum fraction of dark matter captured by PBHs from the background occurs at the conservative upperbound $f_{\textrm{DM}}\simeq 0.01$. In this scenario, the final mass of dark mini-halos accreted around PBHs will decrease $\sim (50-55)\%$ of the original dark matter background at $z\sim 30$~\cite{Mack:2006gz}.

At around $z\sim 30$, the dressed PBHs will begin to interact with non-linear systems to be finally incorporated to galactic halos at $z \sim 6$~\cite{Cooray:2002dia}. It is reasonable to assume that the smooth and
dressed PBHs components of dark matter follow the same spatial distribution when the Galactic halo forms~\footnote{A similar assumption was taken in~\cite{Stref:2016uzb}, where authors analyze a dark matter scenario in the Milky Way halo composed by a smooth background and small scale mini-halos of WIMPS, which are formed at around z $\sim$ 80.}. As galaxies evolve, further isolated or clustered dressed PBHs, as well as dark matter background, may be incorporated to galactic halos. Taking into consideration that clustering of dressed PBHs will enhance the accretion rate of dark matter, we expect that the accretion of the smooth dark matter background from PBHs inside galactic halos should keep going as a continuous process.  We also expect dark mini-halos undergo  different levels of disruption~\footnote{Disrupion of dark matter compact objects (without a central PBH) has been studied for several authors. See, for example,~\cite{2017JETP..125..434D,  Tinyakov:2015cgg} for disruption of axion mini-clusters and~\cite{Zhao:2005mb, Berezinsky:2014wya, Goerdt:2006hp, Schneider:2010jr, Law-Smith:2019sqp} for dark matter clumps.}. The dominant disruption processes acting on dark matter substructures correspond to global tides coming from the mean-field potential of the Milky Way, encounters with stars and tidal interactions during disk crossing. Large dark matter substructures with masses greater than $10^{7}\,M_{\odot}$ would be completely disrupted within a galactic radius of $30\,\textrm{kpc}$~\cite{DOnghia:2009yqa}. By contrast, the fate of smaller and denser dark matter substructures is not totally clear and there is a possibility that a significant part of them survive until today in the solar neighborhood. 

We assume that dressed PBHs after formation ends up in galactic halos with a mini-halo mass given by Eqs.~(\ref{mh},\ref{rh}) as 
%\end{linenumbers}
\begin{equation}
M_{\textrm{halo}}(R_{\textrm{halo}})=3^{3/4}\left( \frac{R_{\textrm{halo}}}{0.019\,\textrm{pc}} \right)^{3/4} \left( \frac{M_{\textrm{PBH}}}{M_{\odot}} \right)^{3/4} M_{\odot}\,.\label{halomass}  
\end{equation} 
%\begin{linenumbers}
For the conservative upperbound $f_{\textrm{DM}} \simeq 0.01$, we should expect a mini-halo mass distribution around the  average  $M_{\textrm{halo}} \sim (50-55) M_{\textrm{PBH}}$. There should have dressed PBHs with smaller and larger halo masses, even with halo masses $M_{\textrm{halo}} \simeq 10^2M_{\textrm{PBH}}$. 
The cuspy internal structure of these mini-halos can be readily derived from Eq.~(\ref{halomass}) as~\footnote{Equation~(\ref{rho}) agrees with Ref.~\cite{Berezinsky:2013fxa} up to a numerical factor of two.}
%\end{linenumbers}
\begin{equation}
\rho_{\textrm{halo}}(r) = \frac{1}{4\pi r^2} \frac{dM_{\textrm{halo}}(r)}{dr}
\simeq 6 \times 10^{-21}\,\frac{\textrm{gr}}{{\textrm{cm}}^{3}}\left( \frac{\textrm{pc}}{r} \right)^{9/4} \left( \frac{M_{\textrm{PBH}}}{10^2\,M_{\odot}}  \right)^{3/4}\,,\label{rho}
\end{equation}
%\begin{linenumbers}
which is valid for $r_{\textrm{min}} \leq r \leq R_{\textrm{halo}}$, where $r_{\textrm{min}} = 8G_NM_{\textrm{PBH}}$. This steep density profile was confirmed by N-body numerical simulations performed in~\cite{Adamek:2019gns}.

To disrupt mini-halos is needed to inject to the system an imput of energy greater than the binding energy, $E_b$. We can estimate this binding energy by substracting to the usual Schwarzschild mass, the energy density integrated in a proper spatial integral with a volume element given by $\sqrt{\gamma}d^3x = [1-2G_N m_{\textrm{halo}}(r)/r]^{-1/2} r^2 dr d\Omega$~\cite{Carroll:2004st}. Thus,
%\end{linenumbers}
\begin{equation}
|E_b| \approx 4 \pi G_N \int_{r_{\textrm{min}}}^{R_{\textrm{halo}}} dr\, r\, \rho_{\textrm{halo}}(r)\, m_{\textrm{halo}}(r)\,,
\label{binding}
\end{equation}
%\begin{linenumbers}
where we have used  $G_N m_{\textrm{halo}}(r)/r \ll 1$ inside the mini-halo. Here, $m_{\textrm{halo}}(r)$ is the Schwarzschild mass of the mini-halo within a radius $r$ from the central PBH. 
In the following analysis, we will refer to the magnitude of the binding energy just as $E_b$.

We will use two Milky Way models performed in~\cite{2011MNRAS.414.2446M, 2017MNRAS.465...76M}.
Both models assume a Navarro, Frenk, and White profile (NFW) for the dark matter halo~\cite{Navarro:1995iw}, a two-component (thin/thick) galactic disk in cylindrical coordinates, and an axisymmetric bulge profile~\footnote{By simplicity, we take the spherical version of the axisymmetric bulge profile by approximating the axial ratio to one.}. In addition, the model in~\cite{2017MNRAS.465...76M}
considers a $H_{\textrm{I}}$ and $H_{\textrm{II}}$ gas disks.   

\subsection{Global tides from the Milky Way Halo}
\label{globaltides}
The high density of dark mini-halos around PBHs offers them protection against tidal stripping. The regime of our interest is $r_{\textrm{tidal}} \geq R_{\textrm{halo}}$. Even though tidal stripping may kick out some particles within the tidal radius, this fraction should be a second-order correction due to these particles would correspond to the tail of the velocity distribution, e.g. to the high-speed population close to the escape speed~\cite{Stref:2016uzb}.  

Since $M_{\textrm{halo}}$ is much less than the Milky Way mass and we are interesting in the regime at which the tidal radius is much less than the radial distance from the dressed PBH to the galactic center, we may apply the distant-tide approximation.  

If we do not assume that the host galaxy is a point mass, the tidal radius is calculated to be~\cite{2008gady.book.....B}
%\end{linenumbers}
\begin{equation}
r_{\textrm{tidal}} = \left(  \frac{M_{\textrm{halo}}(R_{\textrm{halo}})+M_{\textrm{PBH}}}{3 M_{\textrm{MW}}(R_0)} \right)^{1/3} \left( 1 - \frac{1}{3} \frac{d\textrm{ln}M_{\text{MW}}}{d\textrm{ln}R }\Bigr|_{R_0} \right)^{-1/3} R_0\,.\label{rtid}
\end{equation} 
%\begin{linenumbers}
Here $R_{\textrm{0}}$ is the radial distance of the dressed PBH from the center of the host under a circular orbit~\footnote{For a non-circular orbit, $R_0$ can be replaced by the perigalatic distance~\cite{1962AJ.....67..471K}.} and $M_{\text{MW}}(R_0)$ is the total mass of the Milky Way within a radius $R_0$. This mass depends on the global mass density profile. We will use the Milky Way model performed in~\cite{2011MNRAS.414.2446M}.
%\end{linenumbers}

\begin{figure}[t!]
\centering
\includegraphics[scale=0.19]{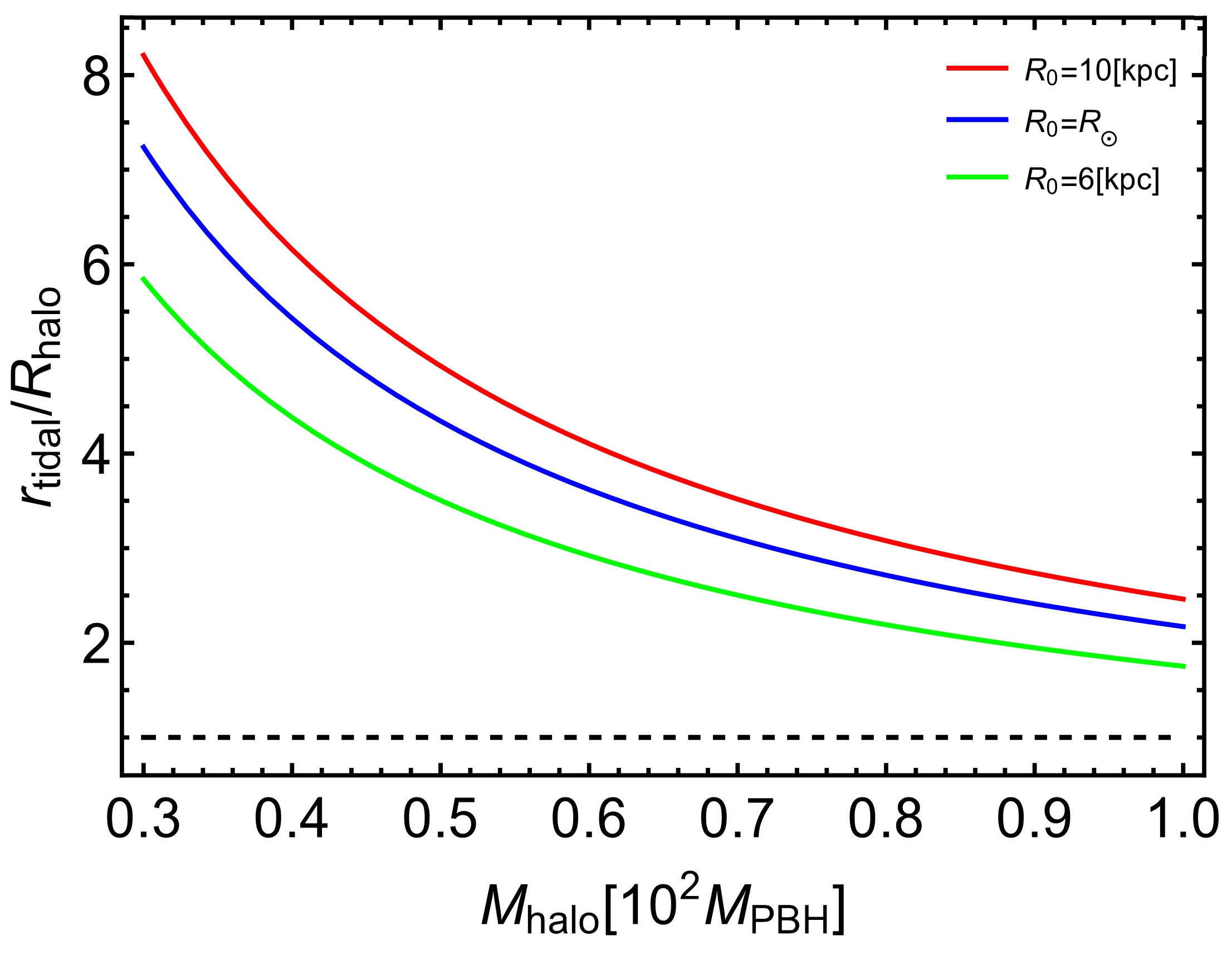}\!\!\!\!\!\!\!\!\!\!\!\!\!\\
\caption{Ratio between the tidal radius and the mini-halo radius, $r_{\textrm{tidal}}/R_{\textrm{halo}}$, with respect to the mini-halo mass, $M_{\textrm{halo}}$, due to global tides from the mean-field potential of the Milky Way. The green, blue and red solid lines show results for $R_0 = (6\,\textrm{kpc}, R_{\odot}, 10\,\textrm{kpc})$ in Eq.~(\ref{rtid}), respectively. The Milky Way mass within a radius $R_0$ is calculated using the Milky Way Model performed in~\cite{2011MNRAS.414.2446M}. The dashed black line at which  $r_{\textrm{tidal}}/R_{\textrm{halo}} =1$ is shown as a reference. }
\label{Tidal}
\end{figure}

%\begin{linenumbers}
The tidal radius runs with the mini-halo mass as $r_{\text{tidal}}/R_{\text{halo}} \sim M_{\text{PBH}}/M_{\text{halo}}(R_{\text{halo}})$. Thus, the larger the mini-halo mass (in units of the central PBH mass), the smaller the tidal radius (in units of the mini-halo radius). You can readily see that from Eq.~(\ref{rtid}) by taking $M_{\text{halo}}(R_{\text{halo}}) \gg M_{\text{PBH}}$ and using Eq.~(\ref{halomass}) to obtain
%\end{linenumbers}
\begin{align}
\hspace{-0.4cm}\frac{r_{\text{tidal}}}{R_{\text{halo}}} \sim 2 &\left(\frac{R_0}{8.29\,\text{kpc}}\right) \left(\frac{10^2\,\text{M}_{\text{PBH}}}{M_{\text{halo}}}\right)   \times\nonumber \\ 
&\hspace{0.7 cm}\left( \frac{10^{11}\,M_{\odot}}{M_{\text{MW}}(R_0)} \right)^{1/3} \left( 1 - \frac{1}{3} \frac{d\textrm{ln}M_{\text{MW}}}{d\textrm{ln}R }\Bigr|_{R_0} \right)^{-1/3}\,. 
\end{align}  
%\begin{linenumbers}
Figure~\ref{Tidal} shows the evolution of the ratio $r_{\textrm{tidal}}/R_{\textrm{halo}}$ at three different radial position at the Milky Way: $R_0 = (6, R_{\odot},10)$. 
In the mass range of our interest, we see mini-halos are resistant in the solar neighborhood against tidal stripping coming from the mean-field potential of the Milky Way. We find tidal stripping becomes sizeable at $R_0 \lesssim (0.6 - 2.5)\, \textrm{kpc}$, e.g. $r_{\textrm{tidal}}/R_{\textrm{halo}} \lesssim 1$, for mini-halos with $M_{\text{halo}} \simeq  (50- 10^2)\, M_{\text{PBH}}$, respectively. 
%\end{linenumbers}
\begin{table*}[hbt]
\begin{center}
\caption{Best-fit models for the Milky Way Galaxy performed in~\cite{2011MNRAS.414.2446M, 2017MNRAS.465...76M}. These models use a NFW halo profile, $\rho_{\textrm{halo}} = \rho_s (r/r_s)^{-1} (1+r/r_s)^{-2}$ and a  two-component disk,  $\rho_d(R,z) = [\Sigma_d/(2 z_d)] \textrm{Exp}(-R/R_d - |z|/z_d)$. The spherical version of the bulge profile reads as $\rho_{\textrm{bulge}}(r) = \rho_b (1+r/r_b)^{-\alpha_b} \textrm{exp}[-(r/r^c_b)^2]$. The gas disk in~\cite{2017MNRAS.465...76M} follows the density profile $\rho_d(R,z) = [\Sigma_d/(4z_d)]\textrm{exp}[-R_m/R - R/R_d]\textrm{sech}^2(z/2z_d)$, where $R_m = (4,12)\,[\textrm{kpc}]$ for $H_{\textrm{I}}$ and $H_{\textrm{II}}$, respectively. Here $r_s$ and $r_b^c$ are in units of kpc, and $\rho_b$ and $\rho_s$ in units of $M_{\odot}/\text{pc}^3$ and $\text{GeV}/\text{cm}^3$, respectively. }
\vspace{0.1cm}
\begin{tabular}{rrrrrrrrrr}
\hline
  MW-model &$r_s$ & $\rho_s$  & $\alpha_b$ &  $r_b$&  $r^c_b$&  $\rho_b$&  $R_d [\textrm{kpc}]\,\,\,\,\,\,\,\,\,\,\,\,\,\,\,\,\,\,$&  $z_d [\textrm{kpc}]\,\,\,\,\,\,\,\,\,\,\,\,\,\,\,$&  $\Sigma_d [M_{\odot}/\textrm{pc}^2]\,\,\,\,\,\,\,\,\,\,$\\   
       & &      &  & &  &  &  $(\textrm{thin}/\textrm{thick})\,(\textrm{HI}/\textrm{HII})$&  $(\textrm{thin}/\textrm{thick})\,(\textrm{HI}/\textrm{HII})$&  $(\textrm{thin}/\textrm{thick})\,(\textrm{HI}/\textrm{HII})$\\          
\hline
Ref.~\cite{2011MNRAS.414.2446M} & $20.2$  & $0.32$   & $1.8$  & $0.075$& $2.1$ & $95.6$& $(2.9/3.31)(-/-)$& $(0.3/0.9)(-/-)$& $(816.6/209.5)(-/-)$ \\
Ref.~\cite{2017MNRAS.465...76M} & $19.6$  & $0.32$  & $1.8$  & $0.075$ & $2.1$ & $98.4$ & $(2.5/3.02)(7/1.5)$ & $(0.3/0.9)(0.085/0.045)$ & $(896/183)(53.1/2180)$\\
\hline
\end{tabular}
\end{center}
\end{table*}

%\begin{linenumbers}
\subsection{Encounters with Stars}
\label{es}
High-speed encounters with stars in the galactic disk, bulge and halo may lead to mini-halo disruption by increasing its internal energy $\Delta E$. For sufficient small impact parameter $b$, one encounter may lead to a total disruption of the dark mini-halo if $\Delta E$ is larger than the binding energy of the system, $E_b$.  The so-called impulse approximation
holds when the typical star-mini-halo encounter is much shorter than the dynamical time scale of the mini-halo. In particular, 
the impulse approximation is valid when the adiabatic parameter $\eta(b) \equiv \omega(b)\tau$ is much less than one, where
$\omega(b)$ is the orbital frequency of the DM at a distance $b$ from the central PBH and $\tau =2 R_{\textrm{halo}}/v_{\textrm{rel}}$ is the typical duration of the encounter. Here $v_{\textrm{rel}} \sim 220\, \textrm{km/s}$ is the typical relative velocity between both astrophysical objects. We can estimate the orbital frequency by considering the inner dispersion velocity as
$\omega(b) = \sqrt{\langle v_{\textrm{DM}}^2(b) \rangle} /b$. By simplicity, we will use the isothermal approximation, where
each Cartesian component of the velocity dispersion is related to the circular velocity as $\sigma^2_{v,i}(b) = \frac{1}{2}v^2_c(b)$. Thus, we obtain
%\end{linenumbers}
\begin{align}
&\tau = 0.026\,\textrm{Myr} \left( \frac{220\,\text{km/s}}{v_{\text{rel}}} \right) \left( \frac{M_{\textrm{halo}}}{10^2\,M_{\textrm{PBH}}} \right)^{4/3} \left( \frac{M_{\textrm{PBH}}}{M_{\odot}} \right)^{1/3}\,,\\
&\omega(b) = 0.55\,\textrm{Myr}^{-1}
 \left( \frac{M_{\textrm{PBH}}}{M_{\odot}} \right)^{3/8}   \sqrt{\frac{(b/\text{pc})^{3/4}-(r_{\textrm{min}}/\text{pc})^{3/4}}{(b/\text{pc})^3}}\,.\label{omega}
\end{align}
%\begin{linenumbers}
In the mass range of our interest, we have $\eta\gg 1$ for orbits close to the mini-halo center, e.g. $r/R_{\textrm{halo}} \lesssim 10^{-3}$. Thus, the impulse approximation only breaks down for small  impact parameter $b$ in units of the mini-halo radius. On the other hand, if the impact parameter is much greater than the mini-halo radius, the star potential will change smoothly across the mini-halo so that the potential may be expanded in a Taylor series and second order terms may be dropped in the distant-tide approximation. When $b \gg R_{\textrm{halo}}$,  the increase of the internal energy of the mini-halo for a single encounter with a star with a mass $M_{\star}$ reads as~\cite{2008gady.book.....B, 1958ApJ...127...17S}
%\end{linenumbers}
\begin{equation}
\Delta E_{\gg} \approx \frac{4 \langle r^2 \rangle}{3}\frac{G_N^2 M_{\star}^2M_{\textrm{halo}}}{v_{\textrm{rel}}^2b^4}\,,\label{bgg}
\end{equation}
%\begin{linenumbers}
where $\langle r^2 \rangle = (1/M_{\textrm{halo}})\int_{r_{\textrm{min}}}^{R_{\textrm{halo}}} d^3r\, r^2 \rho_{\textrm{halo}}(r)$ is the 
mass-weighted mean-square radius of the mini-halo. 

For impact parameters on the order of the mini-halo radius, the distant-tide approximation is no longer valid. In addition, the impulse approximation begins to fail as we approach to the inner parts of the mini-halo.  In the extended regime at which $b \ll R_{\textrm{halo}}$\, or \,$b \gg R_{\textrm{halo}}$, we may use the following parametrization performed in~\cite{1999ApJ...516..195C}:
%\end{linenumbers}
\begin{equation}
\Delta E_{\gg \ll} \approx  \frac{16 \pi}{3}\left( \frac{G_N M_{\star}}{v_{\textrm{rel}} b^2} \right)^2 
\int_{r_{\textrm{min}}}^{R_{\textrm{halo}}} dr\, r^4 \rho_{\textrm{halo}}(r) \left( 1 + \frac{4 r^4}{9 b^4}  \right)\left( 1+ \frac{2r^2}{3b^2}  \right)^{-4}\,.\label{bsoso}
\end{equation} 
%\begin{linenumbers}
Taylor expanding this expression at around $(r/b) =0$, we see that Eq.~(\ref{bsoso}) asymptotically approaches to Eq.~(\ref{bgg}) when $b/R_{\textrm{halo}} \rightarrow \infty$. In particular, for the case of a mini-halo with a mass of $10^2 M_{\textrm{PBH}}$, we have $(\Delta E_{\gg \ll}/\Delta E_{\gg}) \gtrsim 0.98$ for $b/R_{\textrm{halo}}\gtrsim 10$. For smaller values of $b$, Eq.~(\ref{bgg}) 
overestimates the actual change on the internal energy per encounter. Using Eqs.~(\ref{halomass},\ref{rho}), we can explicity calculate the gained internal energy of
the mini-halo $\Delta E_{\gg \ll}$ to obtain
%\end{linenumbers}
\begin{align}
\Delta & E_{\gg \ll} \approx6.5 \times 10^{-21}\,M_{\odot} \left( \frac{M_{\star}}{M_{\odot}} \right)^2 \left( \frac{220\,\textrm{km/s}}{v_{\textrm{rel}}} \right)^2\times  \nonumber\\
&\left(  \frac{R_{\textrm{halo}}}{b} \right)^2 \left( \frac{M_{\textrm{PBH}}}{M_{\odot}} \right)^{1/3} \left( \frac{10^2\,M_{\textrm{PBH}}}{M_{\textrm{halo}}}  \right)^{5/3}\left[ \frac{S(R_{\textrm{halo}}/b)}{S(1)} \right]\,,\label{Esoso}
\end{align}
%\begin{linenumbers}
where $S(1)=0.066$ and
%\end{linenumbers}
\begin{equation}
S(R_{\textrm{halo}}/b) = \sum_{i=1}^{4} C_{i}\, {}_{2}F_{1}  \left[\frac{3}{8},i,\frac{11}{8},-\frac{2}{3}\left( \frac{R_{\textrm{halo}}}{b}\right)^2\right]\,,\nonumber
\end{equation}
%\begin{linenumbers}
with $\left\{C_i\right\}_{i=1}^{4} = (1,-3,4,-2)$ and $_2 F_1$ as the Gauss hypergeometric function. 

To analyze the mini-halo disruption, we need to compare the gained internal energy per single encounter with the binding energy of the system. Using Eqs.~(\ref{rho}, \ref{binding}), we have
%\end{linenumbers}
\begin{equation}
E_b \approx 2.4 \times 10^{-10}\,M_{\odot} \left(  \frac{M_{\textrm{halo}}}{10^2\,M_{\textrm{PBH}}} \right)^{2/3}\left( \frac{M_{\textrm{PBH}}}{M_{\odot}} \right)^{5/3}\,.\label{Eb}
\end{equation}
%\begin{linenumbers}
 We define  the critical impact parameter $b_{\textrm{c}}$ such that 
$\Delta E(b_{\textrm{c}}) = E_b$. All single encounters with an impact parameter $b < b_{\textrm{c}}$ will immediately lead to mini-halo disruption or a significant loss of mass from the mini-halo. This kind of encounter is sometimes called in the literature as one-off disruption. However, the mini-halo disruption may also occur by cumulative effects of several encounters with impact parameters $b \geq b_{\textrm{c}}$. By inspection, we see that the mini-halo binding energy is several order larger than the increase in the internal energy at $b \sim R_{\textrm{halo}}$. Thus, only encounters with very small impact parameter (in units of the mini-halo radius) will lead to one-off disruption. 

To obtain an explicit value for $b_c$, we can numerically solve the trascendental equation $\Delta E_{\gg \ll}(b_c) = E_b$. However, we may equate Eq.~(\ref{Esoso}) to Eq.~(\ref{Eb}),
after expanding $S(R_{\textrm{halo}}/b)$ of Eq.~(\ref{Esoso}) in a Taylor series at around $b/R_{\textrm{halo}}=0$. To leading order, we obtain
%\end{linenumbers}
\begin{equation}
\frac{2b_c}{3R_{\textrm{halo}}} \approx 10^{-8} \left( \frac{M_{\star}}{M_{\odot}} \right)^{8/5} 
 \left( \frac{220\,\frac{km}{s}}{v_{\textrm{rel}}} \right)^{8/5} \left( \frac{M_{\odot}}{M_{\textrm{PBH}}} \right)^{16/15} \left( \frac{10^2\, M_{\textrm{PBH}}}{M_{\textrm{halo}}} \right)^{28/15}\,.\label{bc}
\end{equation}
%\begin{linenumbers}
The larger the ratio $M_{\textrm{halo}}/M_{\textrm{PBH}}$, the smaller the minimal impact parameter (in units of the mini-halo radius) required to lead to one-off disruption of the mini-halo.       

The total probability of disruption for mini-halos when they cross a stellar field, $N_{\textrm{total}}$, is given by the sum of one-off disruptive events, $N_{\textrm{one-off}}$, and multiple encounters with impact parameter $b \geq b_c$, $N_{\textrm{multiple}}$. Since 
the total number of encounters over time $t$ depends on the impact
parameter as $dN = 2\pi n_{\star}v_{\textrm{rel}}\, dt\, b\, db$, we have
%\end{linenumbers}
\begin{equation}
N_{\textrm{total}} 
= \int_0^{b_{\textrm{c}}} dN + \frac{1}{E_b}  \int_{b_c}^{\infty} \Delta E_{\gg \ll}(b)dN  \,\label{pd}\,.
\end{equation}
%\begin{linenumbers}
Due to the distant-tide and impulse approximation do not hold at the length scale of the order of the critical impact parameter, we use Eq.~(\ref{bsoso}) to calculate the gained internal energy for $b \geq b_c$.
The contribution of one-off disruptive events is readily calculated as
%\end{linenumbers}
\begin{equation}
N_{\textrm{one-off}} = \pi n_{\star} v_{\textrm{rel}}\, t\, b_c^2\,,
\end{equation}
%\begin{linenumbers}
where $b_c$ is given by Eq.~(\ref{bc}).
Take a typical star mass $M_{\star} \sim M_{\odot}$ and a  relative velocity $v_{\textrm{rel}} \sim 220\,\textrm{km/s}$.
Since the critical impact parameter is very small in comparison to the mini-halo radius is very unlikely that mini-halos undergo one-off disruption.  In particular, we have $ (n_{\star}/0.1\,\textrm{pc}^{-3})(t/10^{15}\textrm{Myr}) \lesssim N_{\text{one-off}} \lesssim (n_{\star}/0.1\,\textrm{pc}^{-3})(t/10^{14}\textrm{Myr})$ in the whole parameter space  $(M_{\text{halo}},M_{\text{PBH}})$ given by $ 30 M_{\text{PBH}} \leq M_{\textrm{halo}}  \leq 10^2 M_{\textrm{PBH}}$ and $10 M_{\odot} \leq M_{\text{PBH}} \leq 70 M_{\odot}$.    
For cumulative effects of multiple encounters with an impact parameter $b \geq b_c$, the probability of disruption is larger but still negligible. We have  $(n_{\star}/0.1\,\textrm{pc}^{-3}) (t/10^{8}\textrm{Myr}) \lesssim  N_{\textrm{multiple}} \lesssim  (n_{\star}/0.1\,\textrm{pc}^{-3}) (t/10^{7}\textrm{Myr})$ for the same parameter space $(M_{\text{halo}},M_{\text{PBH}})$ as before.

Mini-halo disruption due to encounter with stars occurs when $N_{\textrm{total}} = 1$.  The local stellar number density in the galactic disk, halo, and bulge  is $n_{\star} \sim (0.1,10^{-4},< 10^{-2})\, \textrm{pc}^{-3}$, respectively~\footnote{ The stellar number density of the halo at $r > 3\, \textrm{kpc}$ can be parameterized as $n_{\textrm{halo},\star} = (10^{-4} M_{\odot}\,\textrm{pc}^{-3}/m_{\star}) (R_{\odot}/r)^3$~\cite{Moore:2005uu}. The stellar number density of the bulge at $1\,\textrm{kpc} \leq r \leq 3\,\textrm{kpc}$ is $n_{\textrm{bulge},\star} = (8 M_{\odot}\,\textrm{pc}^{-3}/m_{\star}) \textrm{exp}[-(r/\textrm{kpc})^{1.6}]$~\cite{Launhardt:2002tx}.}. Thus, the disruption of mini-halos due to encounters with stars at the solar neighborhood is negligible.

A opposite situation happens, for example, for neutralino dark matter microhaloes (without a central PBH), which are formed after redshifts $z\approx 100$, due to they have a critical impact parameter much larger than their radius scale. Thus, these microhaloes undergo a significant disruption due to high-speed encounters with stars in the galactic disk~\cite{Goerdt:2006hp}.    

\subsection{Gravitational Field of the Disk}
\label{ds}

Another way to analyze tidal disruption of mini-halos during disk crossings is to focus on the gravitational field of the disk.
As dressed PBHs cross the galactic plane, this gravitational field will exert a compressive gravitational force over them, 
pinching the mini-halos briefly along the normal to the disk plane. The cumulative effect of succesive pinching may eventually disrupt mini-halos. This process in the literature is known as disk shocking and was proposed for the first time in~\cite{1972ApJ...176L..51O} to study the disruption of globular clusters.     

 When dressed PBHs are just about to cross the disk, the disk can be modelled as an infinite slab so that the disk gravitational field is exerted along the axis perpendicular to the disk. Define $Z \equiv Z_{\textrm{0}}+z$ as the height above the galactic disk midplane of a DM particle in the mini-halo, where $Z_{0}(t)$ is the height of the central PBH (or mini-halo center) and $z$ is the height of a DM particle with respect to the mini-halo center. The change in the $z$-component of the velocity of the DM particles is then given by~\cite{2008gady.book.....B} $\dot{v}_z = -4\pi G_N \rho_d(R,Z_0)z$, where 
$R$ is the radial distance of the dressed PBH from the galactic center during the disk crossing, and $\rho_d(R, Z_{0})$ is the density profile of the thin disk.  If the DM orbital time  in the mini-halo is much longer than the disk crossing time, we may apply the impulse approximation to
obtain the gained energy of a mini-halo per unit of DM mass in a single crossing as~\cite{2008gady.book.....B,  Stref:2016uzb}
%\end{linenumbers}
\begin{equation}
\Delta E  = \frac{1}{2} \langle  \Delta v_z^2 \rangle = \frac{1}{2}\Big\langle \left( \int dt \dot{v}_z \right)^2\Big\rangle\,\approx \frac{32\pi^2 G_N^2  \rho^2_d(R,0) z_d^2}{V_z^2}\langle z^2 \rangle\,,\label{DeltaEpot}
\end{equation} 
%\begin{linenumbers}
where we have taken $z$ as approximately constant during disk crossing, $Z_0(t) = V_z t + \textrm{constant}$ with $V_z$ the $Z$-velocity of the dressed PBH, and $\int_{-z_d}^{z_d} dZ_0 \rho_d(R,Z_{0}) \approx 2\,z_d \rho_d (R,0)$ with $z_d$ as the vertical scale height of the thin disk. 

Even though Eq~(\ref{DeltaEpot}) only holds in the impulse approximation, we can extend its validity beyond that regime by considering an adiabatic correction. We closely follow~\cite{Gnedin:1997vp, Stref:2016uzb}, where adiabatic invariance in the context of stellar clusters and dark matter subhalos, respectively, is studied.  So, we introduce the adiabatic correction which depends on the adiabatic parameter $\eta(r)$ as $A(\eta) = \left( 1 + \eta^2 \right)^{-3/2}$,
where $\eta(r) \equiv \omega(r) \tau_{\textrm{cross}}$. The DM orbital frequency is given by Eq.~(\ref{omega}), but now  $\tau_{\textrm{cross}}(R_{\odot}) \approx 0.67\, (H/150\,\textrm{pc})/(V_z/220\, \textrm{km/s})$, where $V_z$ is the dressed PBH velocity perpendicular to the disk plane in the Galactic frame and $H$ is the effective half-height of the disk. 

For mini-halos with $M_{\textrm{halo}} \simeq 10^2\, M_{\textrm{PBH}}$, we have $A(\eta) \gtrsim 0.8$ for $r/R_{\textrm{halo}} \gtrsim 0.3$, e.g. the disk shocking is close to the maximal efficiency ~\footnote{ If an axion particle gives several orbits around the PBH during a disk crossing, angular momentum conservation would protect it of leaving the mini-halo.}.
However, as we approaches to the very center of these mini-halos the impulse approximation becomes to break down and the disk shocking is inefficient. The disk shocking efficiency slowly decreases as mini-halos become lighter. Hence, we extend the validity of Eq~(\ref{DeltaEpot}) as
%\end{linenumbers}
\begin{equation}
\Delta E(r)  \approx \frac{32\pi^2 G_N^2  \rho^2_d(R,0)\,z^2_d\,r^2}{3 V_z^2} A(\eta)\,,\label{DeltaEpot2}\,
\end{equation} 
\begin{figure}[t!]
\centering
\includegraphics[scale=0.187]{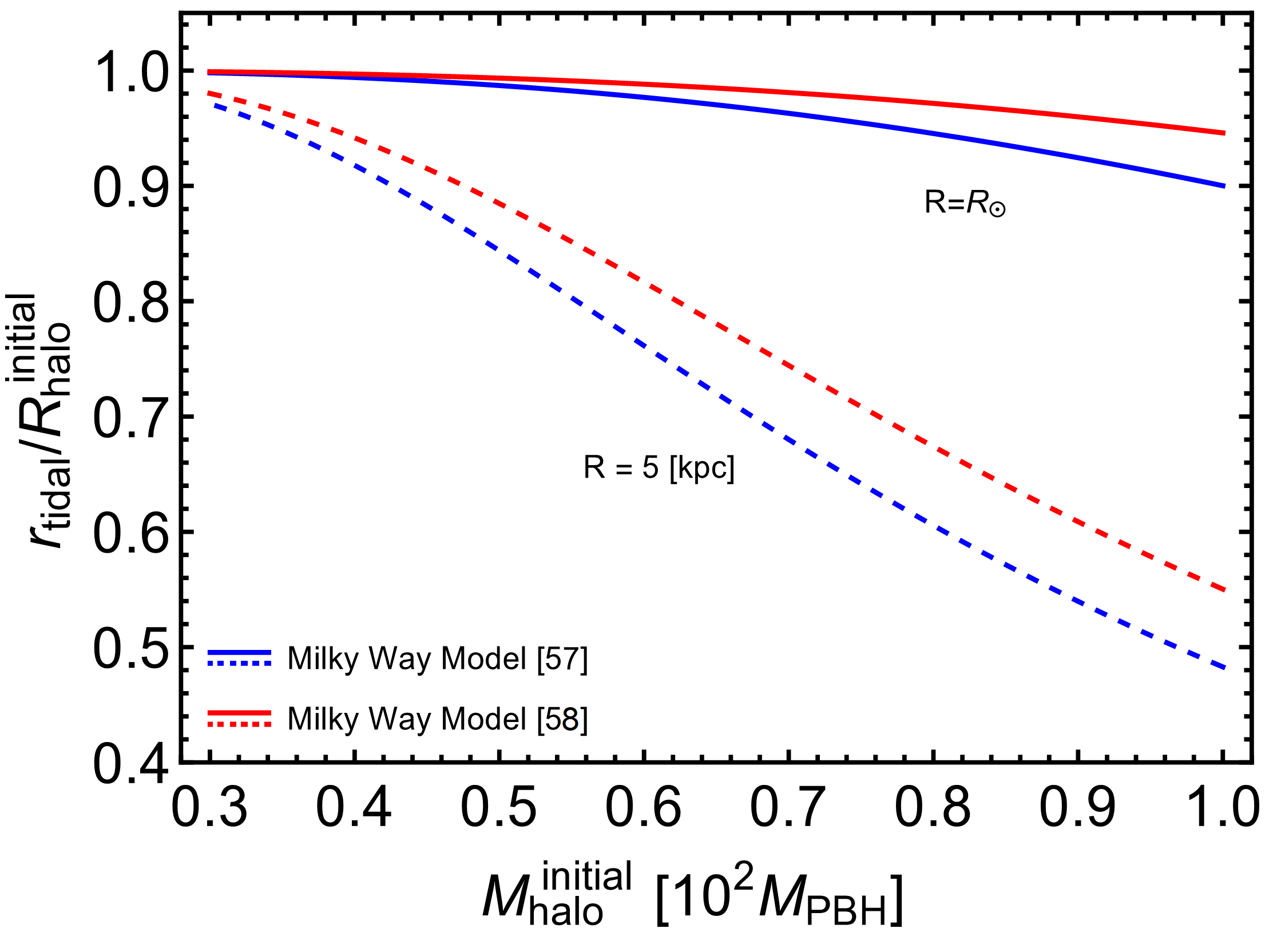}\!\!\!\!\!\!\!\!\!\!\!\!\!\\
\vspace{0.4 cm}
\includegraphics[scale=0.19]{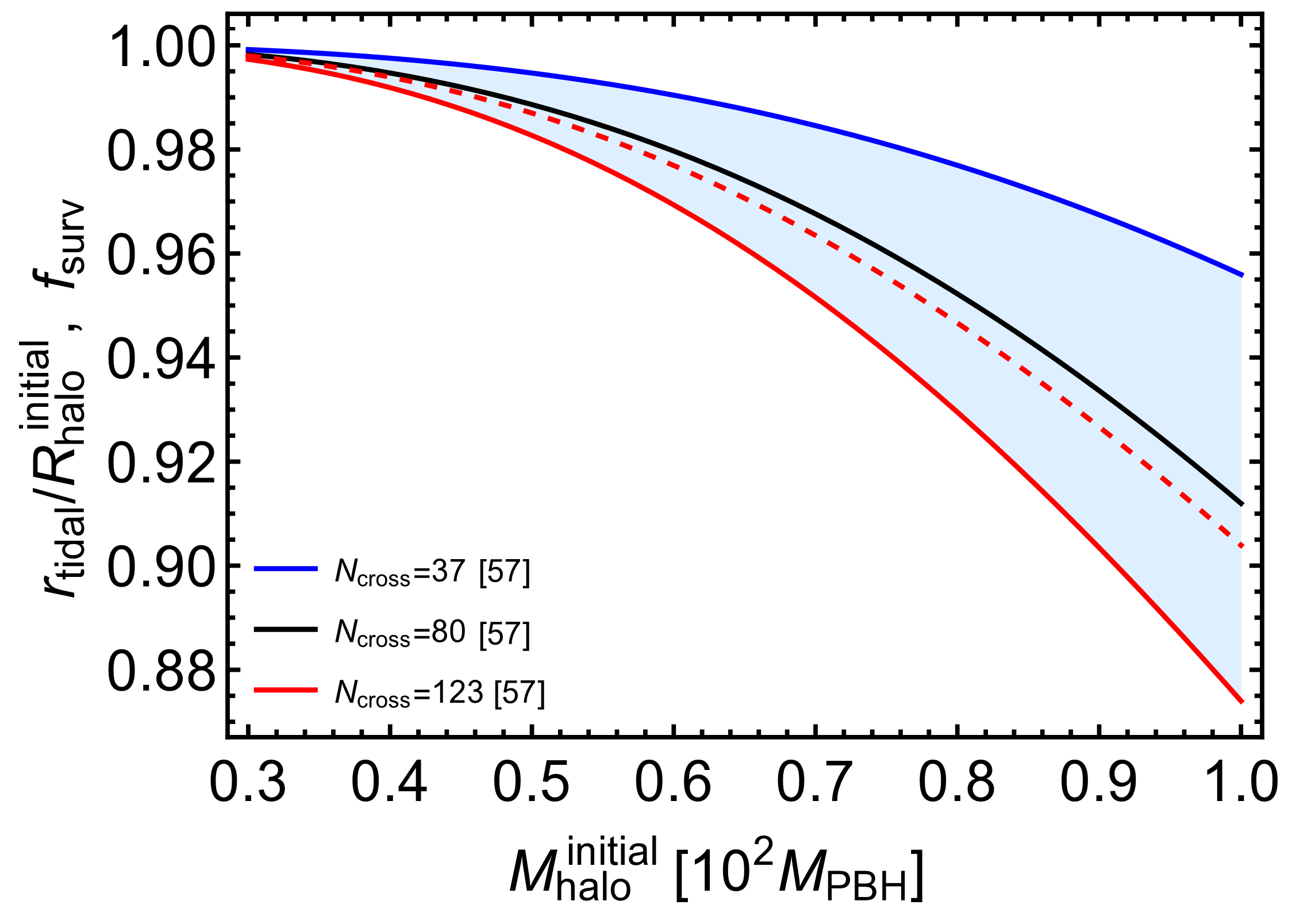}\!\!\!\!\!\!\!\!\!\!\!\!\!\\
\caption{(Top) Ratio of the tidal radius to the initial mini-halo radius, $r_{\textrm{tidal}}/R_{\textrm{halo}}^{\textrm{initial}}$, after disk crossings with respect to initial mini-halo mass, $M_{\textrm{halo}}^{\textrm{initial}}$.  We have used parameters for the thin disk from~\cite{2011MNRAS.414.2446M} (blue lines) and~\cite{2017MNRAS.465...76M} (red lines) assuming circular orbits and $T_{\textrm{MW}}=10\,\textrm{Gyr}$. Solid and dashed lines are calculated at $R = (R_{\odot}, 5\, \textrm{kpc})$, respectively. (Bottom) The same plot as above but using only parameters from~\cite{2011MNRAS.414.2446M} with $N_{\textrm{cross}}=$ (37, 80, 123) at $R=R_{\odot}$ as shown in blue, black, and red solid lines, respectively. The red dashed line shows $f_{\textrm{surv}} \equiv M_{\textrm{halo}}/M_{\textrm{halo}}^{\textrm{initial}}$ with respect to the initial mini-halo mass after $N_{\textrm{cross}} = 123$.}
\label{disrup1}
\end{figure}
%\begin{linenumbers}
\hspace{-0.15 cm}where we have used  $\langle z^2 \rangle = r^2/3$ under the approximation of circular orbits for dark matter particles.

Suppose that a dressed PBH with an initial mini-halo radius $r_{i}$ goes through the galactic disk. We calculate the tidal radius $r_{\textrm{tidal}}$ after crossing by comparing the gained energy of the mini-halo per unit of DM mass with the change in the gravitational potential as follows
%\end{linenumbers}
\begin{equation}
\Delta E (r_{\textrm{tidal}}) = -\left[\phi_{\textrm{halo}}(r_{\textrm{tidal}}) - \phi_{\textrm{halo}}(r_{i})\right]
= G_N \int_{r_{\textrm{tidal}}}^{r_i} dr \frac{m_{\textrm{halo}}(r)}{r^2}\,.\label{iter}
\end{equation} 
%\begin{linenumbers}
We aim to solve Eq.~(\ref{iter}) iteratively so that we can take into account the progressive loss of particles from outer shells to inner shells as the number of disk crossings increases.  We are assuming that between disk crossing dressed PBHs revirialize after a partial loss of particles. The virialization of the mini-halo should be the order of the free-fall time,   $t_{\textrm{free-fall}} = \sqrt{3 \pi / (32 G_N \bar{\rho}_{\textrm{halo}})}$ with $\bar{\rho}_{\textrm{halo}} \sim 3M_{\textrm{halo}}/(4\pi R_{\textrm{halo}}^3)$. Since $t_{\textrm{free-fall}} \simeq (1-8)\, \textrm{Myr}$ for $M_{\textrm{halo}} \simeq (30-10^2)M_{\textrm{PBH}}$, but the half of the circular orbital period for dressed PBHs around the galactic center is $\sim \mathcal{O}(10^2)\,\textrm{Myr}$, our assumption is reasonable. 

Let us suppose that before the first disk crossing a dress PBH has an initial  mini-halo mass and radius given by $M_{\textrm{halo}}^{\textrm{initial}}$ and $R_{\textrm{halo}}^{\textrm{initial}}$, respectively. Figure~\ref{disrup1} (top) shows the corresponding ratios of the final tidal radius to the initial mini-halo radius, $r_{\textrm{tidal}}/R_{\textrm{halo}}^{\textrm{initial}}$, for the Milky Way Models listed in Table I. Results assume circular orbits of dressed PBHs in the galactic frame during the age of the Milky Way, $T_{\textrm{MW}}\sim 10\, \textrm{Gyr}$. Total number of disk crossings
are calculated for each Milky Way model. For example, we have $N_{\textrm{cross}} \sim (155, 93)$ at $R=(5\,\textrm{kpc}, R_{\odot})$ for the model performed in~\cite{2011MNRAS.414.2446M}.  Solid and dashed lines correspond to circular orbits at the sun position and $R = 5\,\textrm{kpc}$, respectively. The smaller the galactocentric radii for disk crossing, the larger the effect of disruption from disk shocking. In addition, similar to the tidal stripping in Sec.~\ref{globaltides}, the level of disruption increases
as the mini-halo mass (in units of the central PBH mass) increases~\footnote{The effect of disk shocking is enhanced by global tides from the mean-field potential of the Milky Way at $R \lesssim (0.6 - 2.5)\,\textrm{kpc}$ for $M_{\text{halo}} = (50 - 10^2)\,M_{\text{PBH}}$. At that regime, before solving Eq.~(\ref{iter}) by iteration, $R^{\textrm{initial}}_{\textrm{halo}}$ needs to be set equal to the tidal radius from global tides as show in Sec.~\ref{globaltides}.}. Here we express the level of disruption  in terms of the ratio of the final mini-halo mass to the initial mass, e.g. $f_{\text{surv}} \equiv M_{\text{halo}}/M_{\text{halo}}^{\text{initial}}$.  The level of disruption over mini-halos at the solar neigborhood is small. For the case of $M_{\textrm{halo}}^{\textrm{initial}} = 10^2M_{\textrm{PBH}}$, the final tidal radius is about $(90-95)\%$ of the initial mini-halo radius. This tidal radius leads to a ratio of the final mini-halo mass to the initial mass of $f_{\textrm{surv}}\simeq (0.92-0.96)$, where $f_{\textrm{surv}} \equiv  (r_{\textrm{tidal}}/R_{\textrm{halo}}^{\textrm{initial}})^{3/4}$ from Eq.~(\ref{halomass}).     

Authors in~\cite{Schneider:2010jr} performed a numerical simulation to study dark matter mini-halo disruption by disk crossing in the solar neighborhood. They used $V_z = \textrm{200 km/s}$, a periodic cosmological box, and the Milky Way model with halo and disk set up by the GalactICS code~\cite{Widrow}. They obtained an average of disk crossings $\bar{N}_{\textrm{cross}}= 80 \pm 43$.  Motivated by that, we recalculate in Fig.~\ref{disrup1} (bottom) the final tidal radius for mini-halos after disk crossings at the solar neighborhood but now using different number of disk crossings.  All results are shown for the Milky Way Model performed in~\cite{2011MNRAS.414.2446M}.   The lightblue shaded region is calculated by taking $N_{\textrm{cross}} = 80 \pm 43$, where the blue, black, and red solid lines correspond to the lower, average, and upper value of $N_{\textrm{cross}}$, respectively. In addition, we show in the red dashed line the fraction of the remaining mass, $f_{\textrm{surv}}$, for mini-halos using $N_{\textrm{cross}} = 123$. We see the loss of mass is still small. In particular, for a mini-halo with an initial mass $10^2 M_{\textrm{PBH}}$, the final tidal radius is up to $87\%$ of the initial mini-halo radius leading  to $f_{\textrm{surv}} \simeq 0.9$. 
 
To calculate with accuracy what fraction of dressed PBHs survive today is needed to consider, through numerical simulations, the angular distribution of their orbits, an accurate model for stellar distribution, dark matter accretion to the mini-halo between disk crossing, radial profile evolution for mini-halos, and gravitational clustering.  We leave this task for future work.

\section{Discussion: Implications for direct detection of DM}

The fraction of dark matter background today in our galactic halo is of crucial importance for direct detection. 
A significant confinement of dark matter in mini-halos around PBHs would affect the flux of dark matter particles on the Earth and, as a result, direct detection experiments would have their sensitivity reduced.  

In the solar neighborhood, we showed in Secs.~\ref{globaltides}-\ref{ds} that dressed PBHs are resistant against tidal forces coming from the mean-field potential of the Milky Way and encounters with stars, but they undergo a small level of disruption coming from the disk gravitational field during disk crossing. Thus, we can focus just on the disruption coming from disk shocking and estimate the local fraction of dark matter in the form of dressed PBHs today as   $f^{\textrm{today}}_{\textrm{halo}} = f_{\textrm{DM}}(f_{\textrm{surv}}M^{\textrm{initial}}_{\textrm{halo}} + M_{\textrm{PBH}})/M_{\textrm{PBH}}$, where the fraction of the remaining mini-halo mass after disk crossings, $f_{\textrm{surv}}$, can be read from Fig.~\ref{disrup1}.  
Accordingly to this, the reduced average of the local dark matter density on the Earth can be estimated as 
%\end{linenumbers}
\begin{equation}
\rho^{\textrm{reduced}}_{\textrm{DM,local}} =
  \rho_{\textrm{DM,local}} \times \left[  1 - \epsilon f_{\textrm{DM}} \frac{(f_{\textrm{surv}}  M_{\textrm{halo}}^{\textrm{initial}}+M_{\textrm{PBH}})}{M_{\textrm{PBH}}} \right]\,,\label{rhoreduced}
\end{equation}
%\begin{linenumbers}
where $\rho_{\textrm{DM,local}}\sim 0.01\, M_{\odot} \textrm{pc}^{-3}$ is the typical local dark matter density, and $\epsilon \lesssim 1$ is a correction factor which encloses uncertainties in our estimates coming from factors not considered by us: capture of dark matter particles to mini-halos from the dressed PBH during their orbits, dressed PBH clustering, and angular distribution for dressed PBHs orbits. These non-trivial features should be studied in future simulations.    

As we mentioned in Sec.~\ref{evolve}, for an initial fraction in PBHs of $f_{\textrm{DM}}\simeq 0.01$,  the Milky Way halo should be populated at the time of formation for 
dressed PBHs with a dominant mini-halo mass about $M_{\textrm{halo}}^{\textrm{initial}} \simeq (50-55) M_{\textrm{PBH}}$.
For this scenario, we see that $f_{\text{surv}} \simeq 0.99$ (0.98) for $M_{\text{halo}}^{\text{initial}} = 50 M_{\text{PBH}}$ ($55 M_{\text{PBH}}$), in Fig~\ref{disrup1}. Thus, the reduced average of the local dark matter density on the Earth is approximately  $\rho_{\textrm{DM,local}} \times[1 - 0.5\,\epsilon]$. Since the disruption coming from disk shocking is proportional to the
ratio $M^{\text{initial}}_{\text{halo}}/M_{\text{PBH}}$, this estimates holds in the whole range of PBH masses of our interest.

As the initial fraction in PBHs decreases, the effect over the local DM background quickly decreases. For example, if $f_{\text{DM}} \simeq 0.001$, the Milky Way Halo should be populated at the time of formation for dressed PBHs with masses around $M_{\text{halo}} \simeq 10^2\,M_{\text{PBH}}$. Thus, the reduced average of the local dark matter density is approximately $\rho_{\text{DM, local}} \times [1 - 0.09\epsilon]$, where we have taken $f_{\text{surv}} \simeq 0.9$ in Eq.~(\ref{rhoreduced}).   
% \end{linenumbers}

% \begin{linenumbers}
Now we want to estimate the possibility that today the Earth goes through the dark mini-halo of a dressed PBH.
 The number of close encounters between the Earth and a dressed PBH per unit of time is given by 
$N_{\oplus\textrm{PBH}_{d}} = n_{\textrm{PBH}_d}\times  \sigma_{\textrm{eff}} v_{\textrm{rel}}$,
where $n_{\textrm{PBH}_d} = f_{\textrm{DM}}\rho_{\textrm{DM,local}}/M_{\textrm{PBH}}$ is the local number density of dressed PBHs, $\sigma_{\textrm{eff}}$ is the usual geometrical cross section enhanced by the gravitational focusing~\cite{2008gady.book.....B}, and $v_{\textrm{rel}} \simeq 220 \,\textrm{km/s}$ is the relative velocity between both astrophysical objects. In particular, the effective cross section can be written as
%\end{linenumbers}
\begin{equation}
\sigma_{\textrm{eff}} = \pi \left[f^{4/3}_{\textrm{surv}}  R^{\textrm{initial}}_{\textrm{halo}}+R_{\oplus}\right]^2 \left[ 1 + \frac{2 G_N (f_{\textrm{surv}}  M^{\textrm{initial}}_{\textrm{halo}}+M_{\oplus})}{v^2_{\textrm{rel}}(f^{4/3}_{\textrm{surv}}  R^{\textrm{initial}}_{\textrm{halo}}+R_{\oplus})} \right]\,.
\end{equation}
%\begin{linenumbers}
In the range of masses of our interest for dressed PBHs, we obtain
%\end{linenumbers}
\begin{equation}
N_{\oplus \textrm{PBH}_d} \sim 0.5 \,\textrm{Myr}^{-1} \left( \frac{f_{\textrm{DM}}}{0.01} \right)  
 \left( \frac{f_{\textrm{surv}}}{0.9} \right)^{8/3} \left( \frac{M^{\textrm{initial}}_{\textrm{halo}}}{10^2\,M_{\textrm{PBH}}} \right)^{8/3}\left( \frac{M_{\odot}}{M_{\textrm{PBH}}} \right)^{1/3} \,.
\end{equation}
%\begin{linenumbers}
This collision rate needs to be compared with the time that the Earth takes to go through a dark mini-halo, e.g. 
%\end{linenumbers}
\begin{equation}
\frac{2 R_{\textrm{halo}}}{v_{\textrm{rel}}}\sim 0.02\,\textrm{Myr}
\left( \frac{f_{\textrm{surv}}}{0.9} \right)^{4/3} \left(\frac{M^{\textrm{initial}}_{\textrm{halo}}}{10^2\,M_{\textrm{PBH}}} \right)^{4/3} \left( \frac{M_{\textrm{PBH}}}{M_{\odot}} \right)^{1/3}\,.
\end{equation}
%\begin{linenumbers}
We may estimate the probability of the Earth  of being inside a dark mini-halo, $P_{w}$, by multiplying this time scale with the collision rate to obtain
%\end{linenumbers}
\begin{equation}
P_{w} \sim 0.01 \times 
 \left( \frac{f_{\textrm{DM}}}{0.01} \right) \left( \frac{f_{\textrm{surv}}}{0.9} \right)^{4} \left( \frac{M^{\textrm{initial}}_{\textrm{halo}}}{10^2\,M_{\textrm{PBH}}} \right)^{4}\,.
\end{equation}
%\begin{linenumbers} 
We have  $0.01\% \gtrsim (100\,P_w) \gtrsim 1\%$ for initial halo masses of $30 M_{\textrm{PBH}} \gtrsim M^{\text{initial}}_{\textrm{halo}} \gtrsim 10^2 M_{\textrm{PBH}}$. Thus, the chance of the Earth being within such an overdense region is small but non negligible. In the unlikely event that it is inside, then there would be a significant increase in the ability of direct detection; however the opposite is somewhat more likely. Developing new experimental probes for this scenario would be useful. 
%\end{linenumbers}

\section{Acknowledgments}
 T. T. Y. is supported in part by the China Grant for Talent Scientific Start-Up Project and the JSPS Grant-in-Aid for Scientific Research No. 16H02176, No. 17H02878, and No. 19H05810 and by World Premier International Research Center Initiative (WPI Initiative), MEXT, Japan. T. T. Y. thanks to Hamamatsu Photonics. M. P. H. is supported in part by National Science Foundation grant PHY-1720332. E. D. S. thanks to Martin Stref for useful discussions.\\



 %\bibliographystyle{elsarticle-num} 
  %\bibliography{WarningPLR_HSY_2020}


\bibliographystyle{elsarticle-num} 
%\bibliography{Warning26042020}
\bibliography{Warning15052020}
%\begin{thebibliography}{00}


%\bibitem{}

%\end{thebibliography}
\end{document}